# USING HAUSDORFF DISTANCE FOR NEW MEDICAL IMAGE ANNOTATION

Riadh BOUSLIMI [1] and Jalel AKAICHI [2]

Department of Computer Sciences
ISG-University of Tunis
Tunisia
[1]bouslimi.riadh@gmail.com
[2]jalel.akaichi@isg.rnu.tn

*ABSTRACT*

*Medical images annotation is most of the time a repetitive hard task. Collecting old similar annotations and assigning them to new medical images may not only enhance the annotation process, but also reduce ambiguity caused by repetitive annotations. The goal of this work is to propose an approach based on Hausdorff distance able to compute similarity between a new medical image and old stored images. User has to choose then one of the similar images and annotations related to the selected one are assigned to the new one.*

*KEYWORDS*

*hausdorff distance, medical image annotation, similarity image, retrieval image, local feature*

## 1. INTRODUCTION

Doctors need to store the medical images annotation in a database to facilitate access and avoid duplication of annotations on the same image of a patient. The first of the current challenges for researchers is the association of semantics to a medical image that has in some cases several pathological. Indeed, the image processing methods are not associating with each image a vector (or vectors) of features computed based on characteristics of images called "low level"(color, texture, etc.) [1]. Querying a database of images is then made by introducing a motion picture in the system and comparing the characteristics and calculated using a similarity measure [2]. No semantics is therefore associated with this process.

One option for assigning semantics to an image is the annotation. The medical images annotation is the task of assigning to each image a keyword or a list of keywords that describe its semantic content. This feature can be seen as a way to make a kind of correspondence between the visual aspects of multimedia data and their low-level features.

The goal of this work is to propose an approach based on *Hausdorff distance* able to compute similarity between a new medical image with old stored images. User has to choose then one of the similar images, and annotations related to the selected one are assigned to the new one. Obviously, we consider that annotation is semi-automatic because it requires user intervention in the selection process.

The remainder of this paper is organized as follows: Section 2 presents a state of the art related to image annotation. Section 3 introduces our annotation approach based on *Hausdorff distance*. To validate our model, we describe our experiments in section 4. We then conclude and give some ideas about our future in work section 5.

## 2. STATE OF THE ART

One challenge for image annotation is to associate semantics to it. There are three types of image annotation: manual, semiautomatic and automatic. For the first type of annotation is done manually by a human responsibility to give each image a set of keywords. The automatic annotation, for its part, is a task performed by a computer and aims to reduce the burden of the user. The first type of annotation increases the accuracy and lower productivity while the second type decreases the precision of the task and increase productivity. To make a compromise between these two tasks, their combination has become necessary. That is what is known as the «semi-automatic annotation».

There are methods that apply a partitioning (clustering) of images and text associated with images thus link the text and image [3].

With this approach, it is possible to predict the label of a new image by calculating probabilities. Minka and Picard [4] have proposed a semi-automatic annotation of images in which the user chooses to annotate regions in an image. Propagation of annotations is done by considering textures. Maron et al. [5] addressed the automatic annotation but using only one keyword at a time. Mori et al. [6] have proposed a "model of co-occurrence" between images and keywords to find the most relevant keywords for an image.

The disadvantage of this model is that it requires a large sample of learning to be effective. Dyugulu et al. [7] proposed a model called "model of translation", which is an improved model of co-occurrence proposed by Mori et al. [6], using a learning algorithm . Probabilistic models such as "Cross Media Relevance Model [1] and "Latent Semantic Analysis" [5] were also proposed. Jia and Wang [8] have used two-dimensional hidden Markov models for annotating images. This work, as we can see, have exploited on the local annotation (using objects).

Several studies on measures of similarity between binary images, the *Hausdorff distance* has often been used in the field of image retrieval by content and has been successfully applied to the matching of objects [1] or in face recognition [9]. For finite point sets, the *Hausdorff distance* can be defined [1] as:

Let $A = \{a_1, ..., a_m\}$ and $B = \{b_1, ..., b_n\}$ denote two finite point sets. Then the *Hausdorff distance* is defined as:

$$H(A, B) = \max(h(A, B), h(B, A)),$$
$$\text{where} \quad h(A, B) = \max_{a \in A} \min_{b \in B} \|a - b\|$$

Hereby h(A,B) is called the directed *Hausdorff distance* from set A to B with some underlying norm $\|.\|$ on the points of A and B.

For image processing applications it has been proven useful to apply a slightly different measure, the (directed) modified *Hausdorff distance*, which was introduced by Dubuisson et al. [10] It is defined as:

$$h_{mod}(A, B) = \frac{1}{|A|} \sum_{a \in A} \min_{b \in B} \|a - b\|$$

We then present a database of medical images and architecture of our system of semi-automatic annotation based on *Hausdorff distance* to assess similarity between images.

## 3. CONTRIBUTION

The manual annotation of medical images with keywords is a gourmet point of view time and so we may have a redundancy of information. As against the automatic annotation lacks both the precision of, for that we will initially present the database to store the annotations (Figure 1). We have implemented it in the DBMS Oracle given its large storage capacity and fast data access.

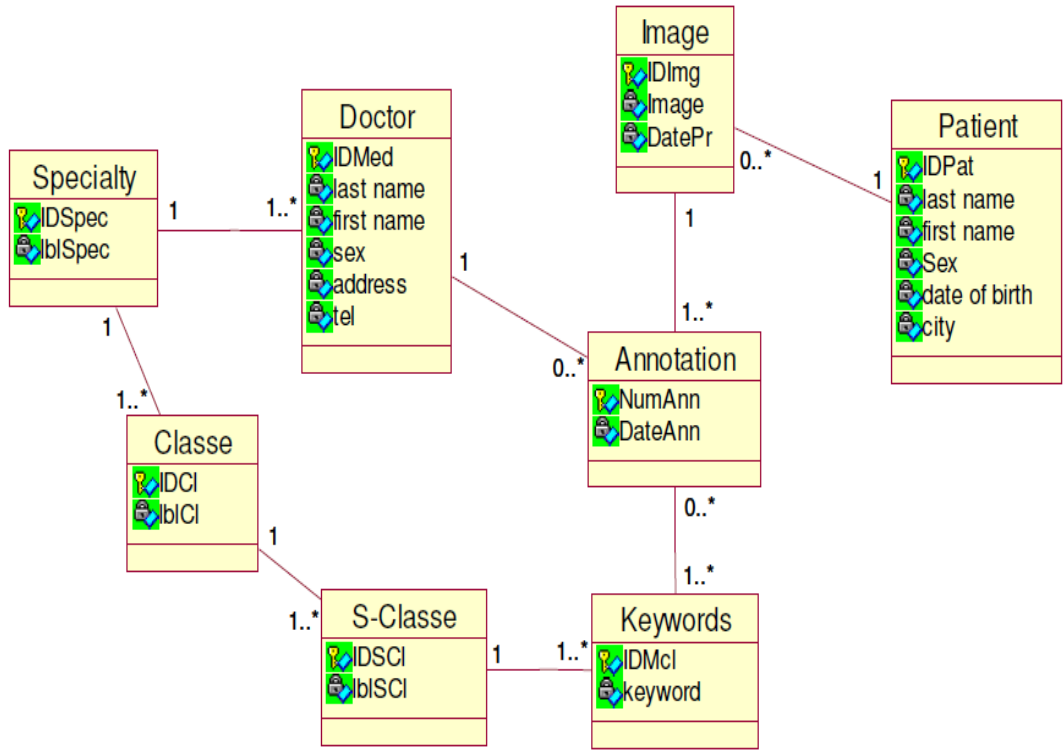

Figure 1: Class diagram of the database of medical images annotation

The power of our database by the annotations on the images is a semi-automatic, meaning that once our system has recognized the image, it will display related keywords to facilitate work for the physician. It may complement other patient information.

Now we will present our system architecture made (Figure 2) based on the combination of the two methods of annotations. In our case (s) doctor (s) (user (s)) is the one who will validate at the end, if the annotation found is good or not. He will record the keywords and other information in the database.

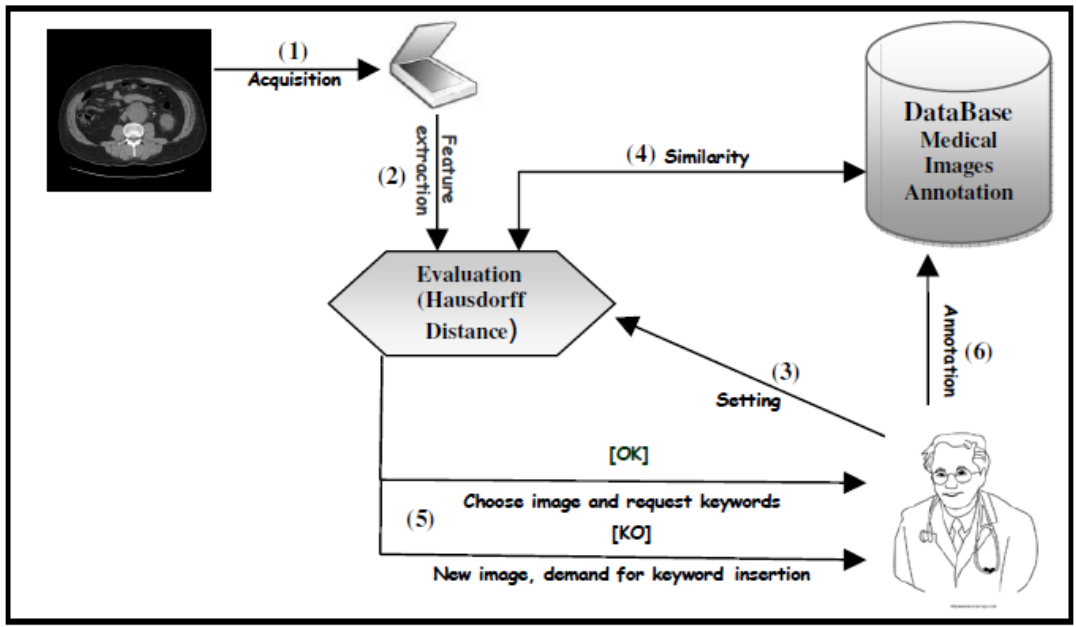

Figure 2: System architecture for semi-automatic annotation of medical images

Figure 2 shows the architecture of the system of semi-automatic annotation of medical images with a final intervention of a doctor responsible for the annotation and validation of keywords found using the *Hausdorff distance* to assess the similarity between images.

We describe explicitly in the following system architecture made: At first, our system starts with a procurement process that is responsible for the cover image, then extract the characteristics of images. After the extraction phase characteristics of the radiographic image, the physician will introduce the specialty and grade of the disease, then it will start the process of similarity and we used to establish the correspondence between the image *Hausdorff distance*. The similarity is made by a round of the database annotation whose specialty fixed in advance by the physician and the class of the disease with the source image. It should be remembered that the comparison of images is widely used in image processing. For binary images that are not composed of simple shapes, a local comparison may be interesting because the extraction of forms is often difficult and the classic attributes (color, texture) poor. Finally, the doctor there could remember which keywords are the most accurate and has the option to validate the annotation of the acquired image, as it can self-select keywords from a fixed listing previously for each specialty to annotate the new image. We will present in the following module similarity.
The annotation can be done several times on a single image, our system allows distinguishing and finding the old annotations of different doctors to ensure accuracy and avoid duplication of images in the database. Figure 3 shows the same image that is duplicated, but it has a variation of the tumor in the colon of a patient. The question here is will we mark the three images in the same way? The answer is yes, the doctor will involve their own keywords, because as you can see, a tumor is present in all images.

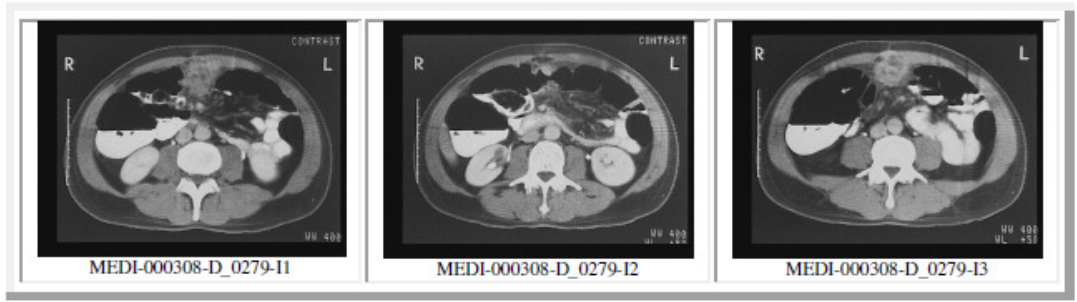

Figure 3: Example of similar medical images

Our database contains annotations for each image the keywords that correspond to it. Here in the table below is an excerpt from our database:

| Classe | Sous-Classe | M-Image | Keywords |
|---|---|---|---|
| | **Speciality : Digestion (1585 images)** | | |
| ABDOMINAL | ABDOMEN | ABD001.jpg | Large abdominalmass, mobile, slow onset. No deterioration of general condition, signsof gastrointestinal or urinary |
| | ABNORMALITIES | ABD002.jpg | Discovered during a routine abdominal ultrasound a right adrenal mass |
| | ABSCESS | ABD003.jpg | Diarrhea for three days. Abdominal pain. Blood count: polymorphonuclear leukocytosis with increased. Small abdominal defense |
| | ANATOMY | ABD004.jpg | feelings of discomfort in the right hypochondrium |
| | ANEURYSMS | ABD005.jpg | Abdominal pain without antecedent pathological |
| | AORTIC | ABD006.jpg | Epigastric pain, a lump supraumbilical fan, aortic aneurysm, in a patient with mucocutaneous jaundice. |
| | AORTOCAVAL | ABD007.jpg | No hypertension, laparotomy with gastric wound suture (penetration motorcycle brake handle). |
| | APPENDICITIS | ABD008.jpg | Diarrhea for three days, abdominal pain, complete blood cell count: leukocytosis with increased neutrophils, Small abdominal defense. |
| | CALCIFICATIONS | ABD009.jpg | Incidental finding of abdominal calcifications |
| | CELIAC DISEASE | ABD010.jpg | malabsorption dominated by diarrhea, steatorrhea. |
| | CHOLECYSTITIS | ABD011.jpg | feelings of discomfort in the right hypochondrium. |
| | COLON | ABD012.jpg | with an abdominopelvic mass of solid nature |
| | CT | ABD013.jpg | Gastric schwannoma, Food intolerance, anorexia, Palpation of an abdominal mass, extrinsic compression of the gastric mucosa in gastroscopy. |
| | ... | ... | ... |

Table 1: Extract the contents of the preliminary annotation DB Medical (Speciality: Digestion)

We achieved a system of treaty radiological images such as "digestive". The latter is based on a user-preset threshold for accepting an endorsement from another. We work here only in setting the decision threshold. We then present an experimental study.

## 4. EXPERIMENTS AND RESULTS

We will show, using a database of images with annotations made by doctors, the interest of the approach we have proposed for the semi-automatic annotation of medical images.

For this we use the database Iconocerf is the digital image bank of Radiology of the Council of France (CERF) and The French Society of Radiology (SFR). This database contains 13,715 images of 3860 with clinical cases.

To perform our experiments, we manually annotated all images with keywords describing the nature of the forms (that is there?). Furthermore, we pre-treat the images of the database by applying some image processing algorithms to extract certain features.

At the end of the pretreatment process, each image is represented by features go through the process of similarity between images using the *Hausdorff distance*.

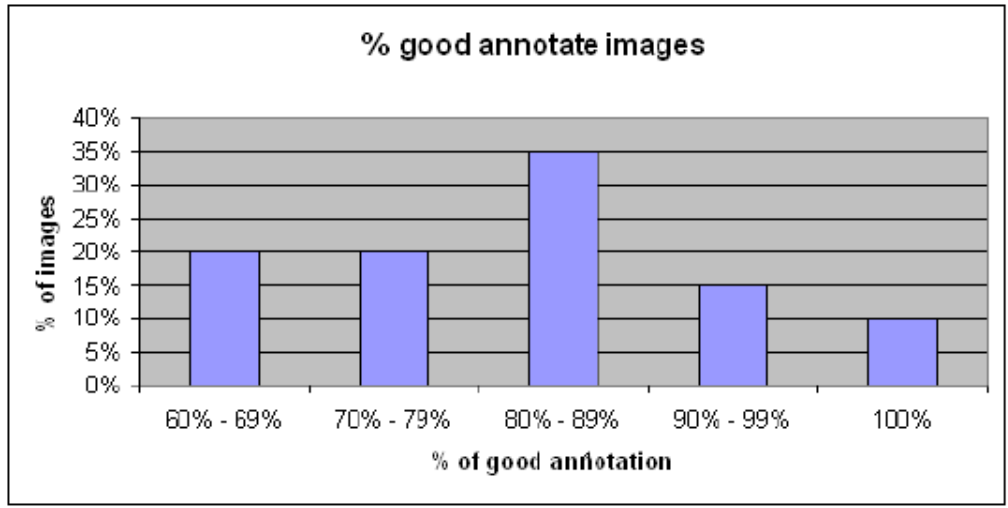

Figure 4 : Statistics of good annotation

The graph in Figure 4 shows the percentage of images by type and by the rate of correct annotation. The first observation is that most images have a good rates of annotation varies between 80% and 90%, which at this stage is a good result annotation.

## 5. CONCLUSION AND PROSPECTS

The annotation of images is the main vehicle for describing semantics to an image. In this article we focus on the semi-automatic annotation of images. Indeed, with the great mass of data managed throughout the world, the manual annotation of these images is virtually impossible. We then proposed a solution based on the method of Hausdorff offering several options: search by content, query by keyword and annotation with voting techniques.

The system we have proposed offers interesting annotation results while meeting the criterion of scalability is a very crucial point in a context where the mass of data is very important.

As perspectives, we plan to first apply our method on a larger database of images more complex than that used in our work and compare our results with results from other methods such as those mentioned in the state of the art. The lack of annotated images prevents us, for now, to make these comparisons. We are also working on integrating knowledge of the user in the annotation process to automate the entire process.